\documentclass[prl,twocolumn,showpacs,preprintnumbers,amsmath,amssymb,floatfix]{revtex4}
\usepackage{epsfig} 
\usepackage{graphicx} 
\usepackage{color}

\newcommand{\ket}[1]{\ensuremath{\vert #1 \rangle}}
\newcommand{\bra}[1]{\ensuremath{\langle #1 \vert}}

\begin{document}

\title{Optimal Control at the Quantum Speed Limit}

\author{T. Caneva$^1$} 
\author{M. Murphy$^2$}
\author{T. Calarco$^{2,3}$} 
\author{R. Fazio$^{4}$}
\author{S. Montangero$^{2,4}$} 
\author{V. Giovannetti$^{4}$}
\author{G. E. Santoro$^{1,5,6}$}

\affiliation{$^1$International School for Advanced Studies (SISSA),
Via Beirut 2-4, I-34014 Trieste, Italy\\ $^2$Institut f\"ur
Quanteninformationsverarbeitung, Universit\"at Ulm, D-89069 Ulm,
Germany\\ $^3$European Center for Theoretical Studies (ECT$^*$), 38050 Villazzano (TN), Italy\\ 
$^4$NEST-CNR-INFM $\&$ Scuola Normale Superiore, Piazza
dei Cavalieri 7, I-56126 Pisa, Italy\\ $^5$CNR-INFM Democritos
National Simulation Center, Via Beirut 2-4, I-34014 Trieste, Italy\\
$^6$International Centre for Theoretical Physics (ICTP), P.O.Box 586,
I-34014 Trieste, Italy}

\date{\today}
\begin{abstract} Optimal control theory is a promising candidate for a
drastic improvement of the performance of quantum information
tasks. We explore its ultimate limit in paradigmatic cases, and
demonstrate that it coincides with the maximum speed limit allowed by
quantum evolution.
% (the quantum speed limit).
\end{abstract}
\pacs{ }
\maketitle

Engineering a suitable Hamiltonian that evolves a given quantum system
into a selected target state 
%(a central issue in many fields of physics) 
has acquired special relevance after the recent advent of
quantum information science~\cite{Nielsen_Chuang:book}. 
Here, the 
%real
challenge is to perform quantum tasks
(e.g. apply a quantum gate)
in an {\em accurate} way
while fulfilling  
%in order to fulfil
 the stringent requirements of fault tolerance.
In this context quantum Optimal Control (OC) is considered a very promising tool,
%to achieve this goal 
and different algorithms have been designed with this aim~\cite{Krotov:book,khaneja05}.
One of them
%way for implementing optimal control is by  exploiting 
exploits 
the Krotov
algorithm~\cite{Krotov:book}, a numerical recursive method which seeks
the OC pulses necessary to implement the required 
%quantum
transformation by solving a Lagrange multiplier
problem~\cite{Peirce_PRA88}. This
technique has already been applied with success to a wide range of
quantum systems~\cite{tannor1,Peirce_PRA88}. One issue
that is not yet fully understood 
however
%about this optimization 
is what its
limits are, and how these limits may be approached. In this work we
will show that the effectiveness of the Krotov algorithm for quantum
OC  is related to fundamental bounds that affect the
maximum {\em speed} at which a quantum system can evolve in its
Hilbert space. Besides being of interest from a theoretical
perspective, the discovery of such a constraint is also  important for
practical implementations. % of optimal control.

For 
%quantum systems driven by generic
 time-dependent Hamiltonians,
bounds that relate the transition probabilities of a quantum system to 
its 
%the system's  
mean energy spread were set by Pfeifer~\cite{Pfeifer_PRL93} and Bhattacharyya~\cite{bhattacharyya}, more than
fifteen years ago.
%, building upon previous results by. 
For time-independent Hamiltonians,
these results have been extended to include dynamical constraints  that
involve also the energy expectation value of the evolving
system~\cite{Margolus_PD98}. 
%{\bf 
Moreover in a specific case 
Khaneja {\em et al.}  evaluated the minimum time required to implement 
a given quantum transformation~\cite{khaneja05}.
In the light of these results  our aim is to 
explore the very limit of  OC.
In particular, we are interested to 
see whether the Krotov algorithm~\cite{Krotov:book}
allows one to attain the ultimate  bound  set by quantum mechanics (for which we
borrow from~\cite{Margolus_PD98} the term Quantum Speed Limit
(QSL)). %.
% on the dynamics of a
%system, or if some other limitations to its effectiveness exist.
 
Several attempts to reconcile accuracy and speed in quantum control
have been proposed so far (see \cite{Carlini_PRL06,Gruebele_PRL07} and references
therein). In particular, Carlini {\em et al.}~\cite{Carlini_PRL06}
cast the time-OC problem into the commonly termed quantum
brachistochrone problem: exploiting the variational principle they
produce a collection of coupled non-linear equations whose solution
(when it exists) yields the required optimal time-dependent
Hamiltonian that minimizes the time evolution while satisfying certain
constraints on the available resources. Our approach differs from that
% GS: Below: in Ref. --> of Ref.
of Ref.~\cite{Carlini_PRL06} since we do not treat the duration of the
process as a variable that enters in the optimization process.
Instead, we set it to some fixed value $T$ and use standard quantum
control optimization techniques to find the $T$-long pulses which
guarantee higher accuracy. The connection between OC and
% GS: Added a T<T_QSL for clarity
the QSL emerges at the time duration $T<T_{\mathrm{QSL}}$ 
for which OC fails to converge.

Specifically, given an input state $|\psi(0)\rangle$ and a Hamiltonian
$H(t)$ that depends on the set of time-dependent control functions
% GS: I eliminated the fancy := If strongly required by someone, please reinsert
${\boldsymbol x}(t) = \{{x}_1(t),x_2(t), \cdots, x_k(t)\}$, we shall
employ the Krotov algorithm~\cite{Krotov:book} to determine the
optimal ${\boldsymbol x}_{opt}(t)$ that minimizes the infidelity
% GS: I eliminated the fancy := If strongly required by someone, please reinsert
$\mathcal{I}=1-|\langle\psi (T)|\psi_G \rangle|^2$, which
measures the distance between the target state $|\psi_G \rangle$ and
% GS: Slightly modified below
the time-$T$ evolved state $|\psi(T) \rangle$ %of $|\psi(0)\rangle$ 
under $H(t)$. The ${\boldsymbol x}_{opt}(t)$ are constructed iteratively,
starting from some initial guess functions ${\boldsymbol x}_{gs}(t)$.
We then analyze the performance of the process as a function of $T$
and show that the method is able to produce  infidelities
%values of ${\cal I}$
arbitrarily close to zero only above a certain threshold
$T_{\mathrm{QSL}}$, which we compare with the dynamical bounds that
affect the system. 
%%%% Tommaso %%%
%Quite surprisingly,
%%%%%%%%%%%%%%%%% 
We found a good agreement between these (in principle) independent 
quantities, meaning that the effectiveness of our control pulses is only limited by the dynamical
bounds of the system. 
%(at least for the models we have explicitly analyzed). 
Considering the limited set of controls we allow in the
problem, and the fact that our initial equations are not meant to
optimize $T$, this is a rather remarkable fact that suggests that OC is a possible candidate for an operational characterization of
the QSLs of complex systems.

Even though our findings have been obtained in several different contexts, including for instance 
ordered Ising and Lipkin-Meshkov-Glick models, 
for the sake of clarity and for demonstration of the generality of the
argument, in this Letter we shall focus on two paradigmatic examples: the
Landau-Zener (LZ) model~\cite{Zener_PRS32}, and the transfer of
information along a chain of coupled spins with Heisenberg
interactions. 
 The former case constitutes a basic step for the
control of complex many-body systems, whose evolution, for finite size
systems, is in many cases a \emph{cascade} of LZ
transitions~\cite{Santoro_SCI02}. Adiabatic quantum
computation~\cite{Farhi_SCI01} is known to be limited by avoided
crossings in the time-dependent system Hamiltonian and by our inability
to avoid excitation of the system. The spin-chain case is instead
related to one of the central requirements for the construction of
circuit-model quantum computers: an infrastructure that can rapidly
and accurately 
%robustly
 transport qubit states between sites~\cite{bose-2002}.
 
%-------------------------------------------------------------------------------
\begin{figure}[t]
\epsfig{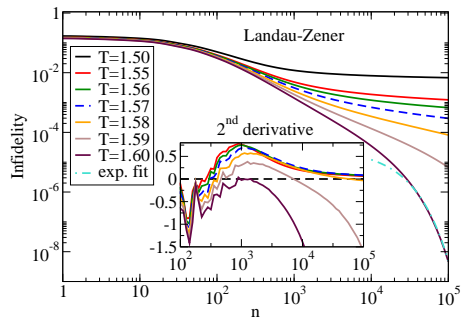}
\caption{(Color online) Infidelity ${\cal I}$
%%%%%%%% Tommaso %%%%%%%%%%%%%%
%${\cal I}= 1 - |\langle \Psi(T) |\psi_G\rangle|^2$ 
%%%%%%%%%%%%%%%%%%%%%%%%%%%%%%%
versus number of iterations $n$ of the Krotov
algorithm~\cite{Krotov:book} for different values of $T$ (in units of $\hbar/\omega$) for $\Gamma
(T)/\omega=500$.
The dashed line 
%%%% Tommaso
%corresponds to
%%%%%%%%%%%%%%%%% 
is the estimated QSL ($T_{\mathrm{QSL}}=1.5688$) 
while the dot-dashed line is an exponential fit.  
Inset: Second derivative of the infidelity logarithm
with respect to the logarithm of $n$.}
\label{iter_vs_infid_lz_fig}
\end{figure}
%-------------------------------------------------------------------------------

{\em Landau-Zener model} - 
The first example we consider is the paradigmatic case of 
the passage through an avoided level crossing
\begin{eqnarray} H[\Gamma (t)]=\small{\left(
\begin{array}{cc} \Gamma (t) & \omega\\ \omega & -\Gamma (t)
\end{array} \right)}\;,
\label{lz_ham_eq}
\end{eqnarray}
in which $\Gamma (t)$ is the control function that we shall optimize
through the Krotov algorithm. 
%In particular, we seek to optimize the
%time taken for passing through the avoided crossing while starting and
%ending in the instantaneous ground state. 
%%% Tommaso
%More precisely,
%%%%%%%%%%%%%%%%%%%% 
We start the evolution by preparing the system in the
instantaneous ground state of $H[\Gamma (0)]$ and we assume as our
target the ground state of $H[\Gamma (T)]$, with $\Gamma(T)=-\Gamma(0)$ 
%%%%%%% Tommaso
%($|\psi(0)\rangle$ and $|{\psi_G}\rangle$ are thus orthogonal
%only in the limit $|\Gamma (T)|=|\Gamma (0)|=\infty$). 
%%%%%%%%%%%%%
($\lim _{|\Gamma (0)|\rightarrow\infty}\langle\psi(0)|{\psi_G}\rangle=0$).
As an initial guess
$\Gamma_{gs}(t)$ for the control we follow
Ref.~\cite{Roland_PRA02}.
Here  on the basis of the adiabatic
theorem~\cite{Messiah:book} the control pulse
$\Gamma(t)$ was selected through a
differential equation $\dot{\Gamma} =\gamma \; G^2(\Gamma)$, where
$G(\Gamma)= 2 \sqrt{ \omega^2 + \Gamma^2}$ is the instantaneous energy
gap of the Hamiltonian~(\ref{lz_ham_eq}), while 
$\gamma = [\arctan(\Gamma (T)/\omega)-\arctan(\Gamma (0)/\omega)]/4T\omega$. 
Starting from the $\Gamma_{gs}(t)$ defined above, we run the OC algorithm for various values of the total time $T$. The
results are reported in Fig.~\ref{iter_vs_infid_lz_fig} by plotting
the infidelity ${\cal I}$ as a function of the iterations $n$ of the
algorithm. 
%%% Tommaso
%As can be clearly seen from the figure,
%%%%%%%%%%%%%%%%%%%%%%%%%%%%%%% 
When $T< T_{\mathrm{QSL}}\approx 1.5688$,
the infidelity ${\cal I}$ does not converge to zero, 
%%%%%%% Tommaso
%the infidelity curvature becomes
%%%%%%%%%%%%%%%
being its curvature asymptotically flat. On the contrary, by progressively increasing $T$
towards and above $T_{\mathrm{QSL}}$, the curvature changes sign and the
infidelity in the large iteration limit decreases
\emph{exponentially}, as confirmed by the fit in
Fig.~\ref{iter_vs_infid_lz_fig}. In the inset of
Fig.~\ref{iter_vs_infid_lz_fig}, data for the second derivative of the
infidelity logarithm with respect to the logarithm 
%of the number of iterations 
of $n$ for different $T$ are shown: the derivative starts to
cross the zero line for $T\approx 1.58$, and for $T> T_{\mathrm{QSL}}$ it clearly
becomes negative. 
%% Tommaso
%It is worth noticing that the change takes place in
%a restricted range of time corresponding to about $2.5\% $ of the
%total evolution time).
%%%%%%%%%%%%%%%%%%%%%%%
These findings are reflected by 
%%% Tommaso
%some interesting features that emerge from 
%%%%%%%%%%%%%%%%%%%%%%%%%%%%%%
the study of the pulse shape of the
optimization process (data not shown). 
For $T < T_{\mathrm{QSL}}$, the
pulse develops a peak which grows indefinitely %in height 
by increasing $n$ and 
%%% Tommaso
%the number of iterations $n$ of the optimization procedure: that is,
%%%%%%%%%%%%%%%%%%%%%%% 
the
control seems unable to converge towards an optimal shape. On the
contrary, when $T > T_{\mathrm{QSL}}$, after a certain number of iterations,
%%% Tommaso
%corresponding approximately to the appearance of the exponential decay regime, 
%%%%%%%%%%%%%%%%%%%%%%%%%%
the shape becomes stable, and only
small corrections of the order of the infidelity take place. 
Remarkably, the peculiar feature of the initial guess $\Gamma_{gs}(t)$
of being almost constantly zero for most of the central part of the evolution
is preserved by the recursive optimization of OC~\cite{Krotov:book}, 
suggesting that for this simple model an estimation of a finite resource QSL bound
$T_{\mathrm{QSL}}$ for $T$ can be deduced by a time-independent
formula, assuming $H_0 =H[\Gamma = 0]$ as the Hamiltonian. In other
words, for most of the evolution time the dynamics can be effectively
described by a time-independent Hamiltonian, which we can use to
analytically estimate the QSL. 
%For transformations induced by
%time-independent Hamiltonians, 
This can be quantified with the
Bhattacharyya bound~\cite{bhattacharyya}, yielding
\begin{eqnarray}\label{www1www} T_{\mathrm{QSL}} \simeq \Delta
E_0^{-1} \arccos |\langle\psi(0)|\psi_G \rangle| \;,
\end{eqnarray}
%
%-------------------------------------------------------------------------------
\begin{figure}[t]
\epsfig{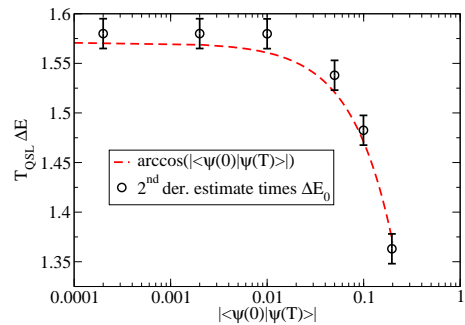}
\caption{(Color online) Comparison between the time independent
estimate (dashed line) and the second derivative criterion (circles,
for $-\Gamma(0)/\omega=5,10,20,100,500,5000$ from right to left) for $T_{\mathrm{QSL}}$ for the LZ model.}
\label{tqslDE_vs_absfid_fig}
\end{figure}
%-------------------------------------------------------------------------------
%
where $\Delta E_0$ is the energy variance of $H_0$ calculated on the
initial state $|\psi(0)\rangle$, i.e. $\Delta E_0 = [ \omega^2 - 4
\omega^4/G^2(\Gamma(0))]^{1/2}$. 
This approach 
has the advantage of providing a bound for $T$ that is
{\em independent} of the effective shape of the selected pulse.
%{\bf GS: I do not understand the meaning of ``not formally correct''.
%It looks like a ``zappa sui piedi''. Do we need this remark?
%It is an estimate, which seems to work rather well. What is the problem with it?}
% VG: sentence modified by removing the "zappa sui piedi"-effect.
Finally,
in Fig.~\ref{tqslDE_vs_absfid_fig} we show a comparison between the
estimate $T_{\mathrm{QSL}}$ through the second derivative of the
infidelity and the theoretical time-independent estimate~(\ref{www1www})
for various $\Gamma (0)/\omega$ ratios. 
We stress that here we have no fitting parameters. 
The excellent agreement  shows that the
OC efficiency is ultimately set by the dynamical bound of
Eq.~(\ref{www1www}).

%{\bf GS: This second part on the Quantum transfer needs still some refinement in the presentation
%of the results. Just a few remarks here and there. 
%1) In Fig.3(left) I do not see the T corresponding to each curve in a legend..
%2) In fig.3(right) I do not understant the color code scale on the right
%axis and the exact meaning of b ...
%3) The plot of Delta E in fig.4(left) seems to me not particularly reach of
%information, but perhaps I am wrong. If we decide to cut it, we might have some
%space to show some optimal pulse shapes either of the LZ or of the QUantum transfer.
%{\bf GS 4) The story about the $\eta$ is not very clear. Do we just want to argue
%that TQSL is linear in N, ``via la testa and via la coda'' (like for the
%grappa) of the chain, where the system accelerates or decelerates?
%If so, this is a simple concept that should be explained in very transparent terms.
%5) It does not seem that we compare this TSQL, which is supposedly optimal for
%the Transfer problem, with anything else previously done on this problem. Any explicit evidence
%of a brilliant improvement over the existing literature would make the article much stronger.}

{\em Quantum state transfer} - We now apply our analysis to a scheme
for information transfer in a spin chain~\cite{bose-2002}. 
In this context a notion of QSL for spin chains was introduced in
Ref.\cite{yung} where the velocity of the information propagation was
optimized with respect to the constant interactions among spins in the
chain. This is of course quite different from the approach we
introduce here where the couplings are given and the information
transfer is sped up by using properly tailored external pulses. 
The model consists of a one-dimensional Heisenberg spin chain of length $N$
described by the Hamiltonian
%
%\begin{equation}\label{ham} 
$H(t) = -\frac{J}{2}\sum_{n=0}^{N-2}\vec{\sigma}_n\cdot\vec{\sigma}_{n+1} +
\sum_{n=0}^{N-1}\frac{C(t)}{2}\bigl(n-d(t)\bigr)^2\sigma_n^z$,
%\end{equation}
%
where $\vec{\sigma} = (\sigma_x, \sigma_y, \sigma_z)$ are the Pauli
spin matrices, $J$ is the coupling strength between nearest-neighbour
spins, $C(t)$ is the relative strength of an external parabolic
magnetic potential, and $d(t)$ represents the position of the
external potential minimum at time $t$ (in units of $\hbar/J$) along the axis of
propagation~\cite{balachandran}. Due to the conservation of the
$z$ component of the magnetization, we can restrict our analysis to
the sector with a single spin-up only, %of unit magnetization, 
so that a general state is described
by $\ket{\psi(t)} = \sum_{m=0}^{N-1} \alpha_m \ket{m}$, where
%$\ket{m}=\ket{\downarrow_0 \cdots \downarrow_{m-1} \uparrow_m \downarrow_{m+1} \cdots \downarrow_{N-1}}$ 
$\ket{m}$ represents the state where the $m$-th site  has its spin pointing up, and all other sites have spins
pointing down. The states $\{\ket{m}\}_{m=0}^{N-1}$ form a complete
orthonormal basis for our Hilbert subspace. Our goal is to evolve the
initial state $\ket{\psi(0)} = \ket{0}$ to the final state
$\ket{\psi_G} = \ket{N-1}$, i.e.\@ to transport a spin up state from
the first site to the final site of the chain. In
Ref.~\cite{balachandran} this was achieved by invoking the adiabatic
approximation, which relies on the fact that slowly moving the
parabolic magnetic potential along the chain allows the spin-up to
migrate from the leftmost site to adjacent sites via a
nearest-neighbour swapping, while interactions between sites far from the field minimum are
frozen. Specifically, the transfer was obtained by assuming pulses of
the form $C(t) = C_0$ and $d(t) = t(N-1)/T$, and by working in a
regime of large (ideally infinite) transmission time $T$. In our
approach,  we shall use instead the Krotov method to find the OC functions $C(t)$ and $d(t)$ using pulses similar to those of
Ref.~\cite{balachandran} as an initial guess: this allows
us to shorten the transfer time beyond what is allowed by the
adiabatic regime. 
%Also in this case 
Once again, we optimize the controls for different values
of $T$ in an effort to identify the minimum transfer time allowed by
the selected controlling Hamiltonian. For each selected
$T$ the optimization algorithm stops either after a certain number $n$
of iterations (of the order of $10^5$) or when the infidelity
$\mathcal{I}$ reaches a certain fixed target threshold $\mathcal{I}^*$.
%
%----------------------------------------------------------------------------
\begin{figure}[t] \centering
  \includegraphics{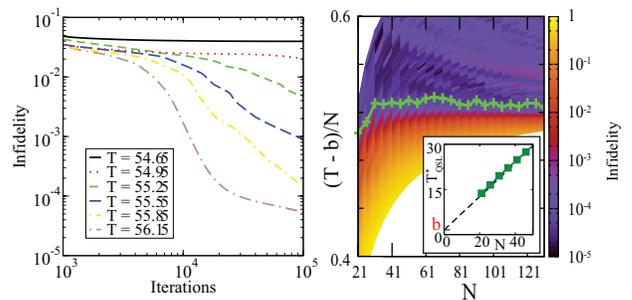}
  \caption{Left: Infidelity as a function of the number of iterations
$n$ 
%%% Tommaso
%of the Krotov algorithm
%%%%%%%%%%%%%%%%%%%%% 
for a spin chain of length $N=101$ and
different durations $T$ (in units of $\hbar/J$). 
%The behavior is qualitatively similar to that
%observed in the Landau-Zener case of Fig.~\ref{iter_vs_infid_lz_fig}. 
%{\bf GS: Ehm... mitigate a bit ...  The curvature is very different ...}
Right: Minimum infidelity
$\mathcal{I}$ for chain length $N$ and the (rescaled) transfer time $(T-b)/N$ 
($b \sim 1.53$ is the $y$-axis intercept of the function
$T^*_{\mathrm{QSL}}(N)$ reported in the inset),
showing the expected linear behaviour of $T_{\mathrm{QSL}}$ with the size $N$.
%{\bf GS: Completely obscure, at this stage, what is the intercept b...}
The green line
% GS: follows the estimate --> follows from the estimate. Is the meaning corrently preserved?
follows from the estimate $T^*_{\mathrm{QSL}}$ of the QSL
time obtained by choosing the time $T(N)$ at which the
infidelity reaches the value $\mathcal{I}^* = 5 \times 10^{-5}$ for $n=10^{5}$.}
  \label{fig:eplot}
\end{figure}
%----------------------------------------------------------------------------
%
The results obtained are shown in Fig.~\ref{fig:eplot} and resemble
those we have seen in the LZ model. In particular, as shown in
Fig.~\ref{fig:eplot} (left) the infidelity appears to converge to
zero only for values of $T$ that are above a certain critical time
$T_{\mathrm{QSL}}$.  On the contrary, for $T<T_{\mathrm{QSL}}$
the convergence of the infidelity slows down, providing numerical
evidence of a non-zero asymptote for $n \rightarrow \infty$. In
contrast with the LZ case however, the dependence of ${\cal I}$ with
the iteration number $n$ is now less regular,
reflecting the fact that the spin-chain dynamics is more complex than
% GS: Changed here a bit ...
for the LZ model. Consequently, for the present model
the sign of the second derivative cannot be used as a reliable
signature of $T_{\mathrm{QSL}}$. 
%{\bf GS: Ops ... Our supposedly general
%scheme does not work on the second example we present ... This should be fixed
%in some way. We should mitigate this very noticeable clash of statements....}
%
Nonetheless, a numerical estimate
$T^*_{\mathrm{QSL}}$ for such quantity can been obtained by considering the smallest
time $T$ which allows us to achieve the target infidelity threshold
$\mathcal{I}^*$ in a fixed number of algorithm iterations $n$ 
(the result does not depend significantly on the
value of $\mathcal{I}^*$ and $n$). 
Apart from the case of small $N$, where boundary
effects are more pronounced, the resulting $T^*_{\mathrm{QSL}}$
appears to have a linear dependence on the chain length $N$ -- see
Fig.~\ref{fig:eplot} (right).

\begin{figure}[t] \centering
  \includegraphics[scale=0.26]{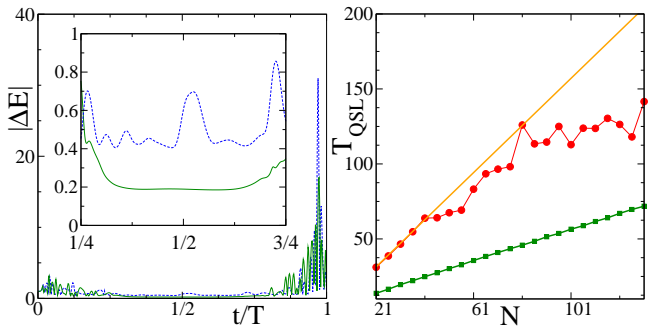}
  \caption{Left: $\Delta E_2(t)$ for a spin-chain of length $N=101$, and
    different total times $T=68.97$ (solid green) and $T=56.66\approx
    T^*_{\mathrm{QSL}}$ (dashed blue) (the times have been rescaled to
    make them fit on the same axis). The central portion of the plot has
    been enlarged in the inset. Right: $\tfrac{\pi (N-1)}{2 \Delta
      \mathcal{E}_1}$ (orange), $\tfrac{\pi (N-1)}{2 \Delta
      \mathcal{E}_2}$ (red circles), and $T^*_{\mathrm{QSL}}$ (green
    squares) versus chain length $N$.  }
  \label{fig:de}
\end{figure}

For a comparison with an independent theoretical estimate of
$T_{\mathrm{QSL}}$, we cannot directly use the 
Bhattacharyya bound (\ref{www1www}), since in this case we are not allowed 
to treat the Hamiltonian as approximately time-independent. 
Nevertheless, a bound on the
minimal transferring time can be obtained by considering the
\emph{mean} energy spread, obtained by averaging the instantaneous
% GS: changed time-independent to time-dependent, below. Is this right
energy spread of the system of the time-dependent Hamiltonian $H(t)$
over the time evolution $[0,T]$. We define this by $\Delta
\mathcal{E}_\lambda = \tfrac{1}{T} \int_0^T \Delta E_\lambda (t)\:
\mathrm{d} t$, 
% GS: Modified below
where $\Delta E_\lambda (t)=\sqrt{\bra{\phi}[H(t) - E_\lambda (t)]^2\ket{\phi}}$
is the energy spread  and $E_\lambda (t) = \bra{\phi}H(t)\ket{\phi}$ 
is computed on the state $\ket{\phi}$, and $\lambda=1,2$ labels
different choices of $\ket{\phi}$. 
For $\lambda = 1$, we follow Ref.~\cite{Pfeifer_PRL93} and take $\ket{\phi}$ to be either the
initial state $\ket{\psi(0)}$ or the target state $\ket{\psi_G}$,
whichever results in the smaller $\Delta \mathcal{E}_1$. 
For $\lambda = 2$, we choose $\ket{\phi} = \ket{\psi(t)}$, which means we
effectively divide the total evolution time $T$ into smaller intervals
$dt$ over which $H(t)$ can be assumed to be constant, and apply to
each of them the Bhattacharyya bound~\cite{bhattacharyya} for
time-independent Hamiltonians. The bound on the minimum transfer time
is then given by
%
%\begin{eqnarray}
%  \label{DEFCORR}
  $ T_{\mathrm{QSL}}/(N-1) \ge \max\{ \tfrac{ \pi }{2 \Delta {\cal E}_1} ;
\tfrac{ \pi }{2 \Delta {\cal E}_2} \}$,
% \;,
%\end{eqnarray}
%
which needs to be satisfied by any $H(t)$ that brings $\ket{\psi(0)}$ to
the target state $\ket{\psi_G}$ in a time $T$. Since we have taken the
time average of the bound, we interpret the quantum speed limit as
describing the minimum transfer time `per-site'. In Fig.~\ref{fig:de}
(left), we report $\Delta E_2(t)$ as a function of time for two
different total transfer times. 
%%%% Tommaso
%As it can be clearly seen,
%%%%%%%%%%%%%%%%%%%%%%%%%% 
The energy
spread is almost constant save near the final time, where large
oscillations are present, corresponding to the deceleration of the
spin-wave. The picture that arises is that OC finds a
solution ${{x}_{opt}(t)}$ that initially accelerates the spin
excitation, and then transfers it with constant velocity up to the end
of the chain, where deceleration occurs.  The average energy spreading
$\Delta E_2(t)$ is larger for smaller time transfer $T$ allowing for a
higher average excitation velocity. We finally compare the estimated
optimal time $T^*_{\mathrm{QSL}}$ from Fig.~\ref{fig:eplot} (right) with
the analytical estimate of the QSL given above.
%by Eq.~(\ref{DEFCORR}). 
The
results are reported in Fig.\ref{fig:de} (right) where the two
quantities are compared: both estimators are linearly dependent on $N$,
but the numerical results show an improvement over the theoretical
prediction by a factor of $\eta \approx 3$, which we attribute to the
difficulty in formulating the quantum speed limit for our many-body
problem. One can think of the optimal transfer of the excitation as being
facilitated by a cascade of \emph{effective} swaps.
%
%-------------------------------------------------------------------------------
%--------------------------------------------------------------------------------
%
As before, the agreement of scaling of the two results shows that even
in the presence of additional constraints the OC reaches
the ultimate dynamical bounds set by quantum mechanics. Indeed, we
achieved an improvement of the transfer time with respect to
Ref.~\cite{balachandran} of up to two orders of magnitude. It is also
worth mentioning that we achieved a transfer time faster than that
obtained (for Ising coupling) in Ref.~\cite{khaneja}. Interestingly
enough, however, our method { only} uses single-site local pulses
while in Ref.~\cite{khaneja} this was achieved
using { global} pulses that operate jointly on the whole chain.

In summary, we have demonstrated that there are fundamental constraints governing the
efficiency of Krotov quantum OC algorithm dictated by the
maximum speed at which a quantum state can evolve in time. These
results provide a further link between control theory and quantum
dynamics.

We thank A. Carlini for discussions, and the bwGRiD (http://www.bw-grid.de)
for the computational resources. We acknowledge financial support by
SFB/TRR21, and the EU under the contracts MRTN-CT-2006-035369 (EMALI),
IP-EUROSQIP, and IP-SCALA.
%
% BIBLIOGRAPHY
%

\bibliographystyle{apsrev} 
%\bibliography{qslandoct}

\end{document}